\documentclass[epj,nopacs]{svjour}

\usepackage{ifthen} % for conditional statements
\newboolean{pdflatex}
\setboolean{pdflatex}{true} % False for eps figures 

\newboolean{articletitles}
\setboolean{articletitles}{true} % False removes titles in references

\newboolean{uprightparticles}
\setboolean{uprightparticles}{false} %True for upright particle symbols
\usepackage{graphicx}
\usepackage{url}
\usepackage{xspace}
\def\geant      {\mbox{\textsc{Geant4}}\xspace}
\def\mitsuba    {\mbox{\textsc{Mitsuba3}}\xspace}
\def\drjit    {\mbox{\textsc{DrJit}}\xspace}
\def\optix    {\mbox{\textsc{OptiX}}\xspace}
\def\freecad    {\mbox{\textsc{FreeCAD}}\xspace}
\usepackage{siunitx}
\usepackage{mciteplus}
\begin{document}

\title{Optical Photon Simulation with Mitsuba3}

\author{Adam C.S. Davis\inst{1}\thanks{ \email{adam.davis@manchester.ac.uk} (Corresponding Author)} \and Sacha Barr\'e \inst{1}\thanks{ \email{sachabarre1711@gmail.com}} \and Yangyang Cui \inst{2}\thanks{ \email{yangyang.cui@student.manchester.ac.uk}} \and Keith L Evans\inst{1}\thanks{ \email{keith.evans@manchester.ac.uk} (Corresponding Author)}
 \and Marco Gersabeck \inst{1}\thanks{ \email{marco.gersabeck@cern.ch}} \and Antonin Rat \inst{1}\thanks{\email{antonin.rat@outlook.fr}} \and
 Zahra Montazeri\inst{2}\thanks{\email{zahra.montazeri@manchester.ac.uk}}
 }

\institute{Department of Physics and Astronomy, The University of Manchester, Oxford Road, Manchester, M13 9PL, United Kingdom \and Department of Computer Science, The University of Manchester, Oxford Road, Manchester, M13 9PL, United Kingdom}
\date{Submitted: 22 September, 2023}
\abstract{Optical photon propagation is an embarrassingly parallel operation, well suited to acceleration on GPU devices. Rendering of images employs similar techniques---for this reason, a pipeline to offload optical photon propagation from \geant to the industry-standard open-source renderer \mitsuba has been devised. With the creation of a dedicated plugin for single point multi-source emission, we find a photon propagation rate of $2\times10^{5}$ photons per second per CPU thread using LLVM and $1.2\times10^{6}$ photons per second per GPU using CUDA. This represents a speed-up of 70 on CPU and 400 on GPU over \geant and is competitive with other similar applications. The potential for further applications is discussed.
\keywords{Particle Physics simulation, Ray Tracing, Cherenkov Radiation, Optical Photon, GPU computing, Multi-architecture, Rendering}
}

\maketitle
%%%%%%%%%%%%%%%%%%%%%%%%%%%%%%%%%%%%%%%%%%%%%%%%%%%
\section{Introduction}\label{sec:introduction}

Propagation of photons produced in High Energy Physics (HEP) simulations is computationally expensive. The propagation of photons is embarrassingly parallel, as each photon can be propagated independently from all others. For this reason, the process is well suited to technologies such as Graphical Processing Units (GPUs), High Performance Computers (HPCs), Cloud deployment or any other emerging technology well suited to embarrassingly parallel tasks. 

To this end, we focus on the exploration of rendering technologies often used in animated movies. The processes employed rely on the same underlying principle---propagation of photons through a scene, which encompasses objects of interest and any other objects within the scene to be rendered, then rendering of optical photons using a camera or detection plane. This paper explores the use of one such renderer, \mitsuba\cite{Mitsuba3}, in the context of particle physics optical photon simulation. We present a prototype workflow for the incorporation of \mitsuba within the \geant\cite{Agostinelli:2002hh,Allison:2006ve} framework, the standard for the interaction of particles with matter in HEP. The workflow is demonstrated in the context of the simulation of a Ring Imaging Cherenkov (RICH) Detector, specifically modelled after those used by the LHCb experiment\cite{Li:2023ocy}. 
These detectors are tasked with the generation and detection of Cherenkov photons, which are emitted by charged particles that traverse a medium at a speed greater than the phase velocity of light in the medium for use in particle identification, and are therefore key ingredients to many experimental apparatuses. 

The paper is organised as follows: Section~\ref{sec:mitsuba_summ} presents a summary of the features of \mitsuba, including the key features necessary to exploit multiple architectures. Section~\ref{sec:cherenkov} discusses the key concepts behind the propagation of Cherenkov radiation within particle physics detectors, including the generation of photons in \geant and the modelling of quantum efficiencies of the detectors. Section~\ref{sec:workflow} presents the prototype workflow on which this work is based, with Section~\ref{sec:geo_translation} dedicated to the translation of geometries from \geant to the \mitsuba renderer, Section~\ref{sec:custom_emitter} defining the implementation of a custom photon emitter in \mitsuba. Section~\ref{sec:results} shows the physics validation of \mitsuba based photon propagation with that of \geant, including a comparison of timing. Section~\ref{sec:discussion} presents an in-depth discussion of the results, and finally conclusions and future work are presented in Section~\ref{sec:conclusion}.

%%%%%%%%%%%%%%%%%%%%%%%%%%%%%%%%%%%%%%%%%%%%%%%%%%%
\section{Summary of Key Features of \mitsuba}
\label{sec:mitsuba_summ}
\subsection{\mitsuba Variants}
\label{sec:variants}
\mitsuba\cite{Mitsuba3} is a physics-based renderer, relying on just-in-time (JIT) compilation through the use of the \drjit\cite{Jakob2020DrJit} engine to generate optimised kernels. \mitsuba natively supports rendering on multiple platforms through the use of so-called variants, which configures the JIT compilation of \mitsuba on which platform to use. Specifically, \mitsuba supports the use of NVIDIA's Compute Unified Device Architecture (CUDA)\cite{cuda} and NVIDIA's \optix\cite{optix} engine for processing on NVIDIA GPUs. Parallel processed ray-tracing on Central Processing Units (CPUs) is supported using LLVM~\cite{LLVM:CGO04} and Intel {\sc Embree}~\cite{Wald:2014:EKF:2601097.2601199}; finally serial operations is enabled via use of the \texttt{scalar} variant, which requires no JIT compilation. As the code supports many different architectures, the design of custom functions following the \mitsuba conventions enables the simultaneous development for multiple different platforms. 

In addition to computational platform, variants are used to select the method by which differing wavelengths are treated during rendering. \mitsuba can represent wavelengths in several modes of interest, \texttt{RGB} and \texttt{mono}. In ray-based rendering tasks, it is often infeasible to represent the visible spectrum of light as a continuum of separately coloured rays. Typically, coloured images are represented by three colour channels, Red, Green and Blue (RGB). In this representation, each channel's value corresponds to the intensity of its respective colour component, e.g. how much red, green or blue is present in the pixel, allowing control over the colour composition of each pixel. Therefore by binning the visible spectrum into low, mid and high frequency ranges, colour images can be generated using three rays per photon. \mitsuba facilitates this wavelength treatment through \texttt{RGB} variants such as \texttt{llvm\_rgb} and \texttt{cuda\_rgb}. The 
\texttt{mono} variants are a simplification of RGB treatment, otherwise known as the intensity treatment, where the wavelength of photons is ignored and the pixel values are represented by a single channel creating a grey scale image. These are selected using the \texttt{mono} variants such as \texttt{llvm\_mono} and \texttt{cuda\_mono}.
%%%%%%%%%%%%%%%%%%%%%%%%%%%%%%%%%%%%%%%%%%%%%%%%%%%
\section{Simulation of Cherenkov Photons---propagation and detection}
\label{sec:cherenkov}
It is well known that when a charged particle passes through a medium at a speed faster than the phase velocity of light within the medium, Cherenkov radiation is produced. The radiation is produced at an angle $\cos\theta_c=1/(n\beta)$ relative to the charged particle's path, where $\beta = v/c$, is the velocity of the particle relative to the speed of light in vacuum and $n$ is the refractive index of the medium. The angle $\theta_c$ is independent of azimuthal angle with respect to the charged particle's velocity. When combined with momentum measurements, the reconstruction of the angle of Cherenkov radiation originating from a charged track allows for particle identification. The \geant toolkit is able to produce Cherenkov radiation; For this reason, we do not explore the generation of Cherenkov Radiation by \mitsuba itself, but rather treat the emitted Cherenkov photons as input to the ray tracing provided by \mitsuba. 

Detection of Cherenkov photons is normally performed by either photomultipliers (PMTs), or similar technologies, and may include the reflection of the photons from a set of mirrors or surfaces. As these technologies also have a response that is dependent on photon wavelength, the modelling of the wavelength of photons is extremely important.
%%%%%%%%%%%%%%%%%%%%%%%%%%%%%%%%%%%%%%%%%%%%%%%%%%%
\section{Prototype workflow}
\label{sec:workflow}
\begin{figure}[!htb]
    \centering
    \includegraphics[width=0.5\textwidth]{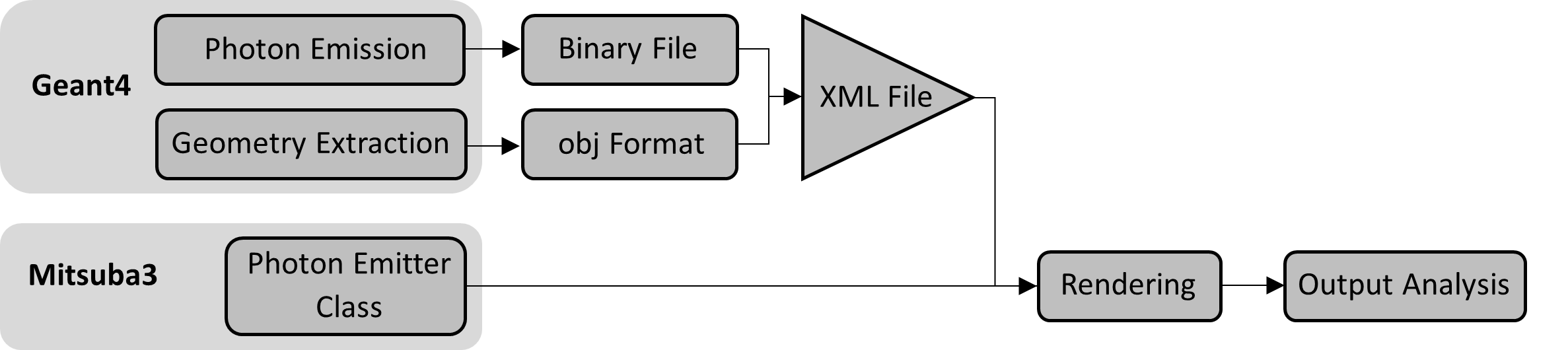}
    \caption{Prototype workflow of the incorporation of 
    \mitsuba within the \geant framework. Each step was designed to be replaceable by a memory-resident implementation.}
    \label{fig:workflow}
\end{figure}

A prototype workflow was designed to aid in the transition to a fully-fledged \geant workflow and to enable simultaneous developments at all levels. This is illustrated in Figure~\ref{fig:workflow}. Each part of the workflow was designed with a memory-resident intermediate implementation in mind, allowing for intermediate file writing to be removed when each step is finished. We enumerate the individual elements of the workflow in the subsequent sections.

%%%%%%%%%%%%%%%%%%%%%%%%%%%%%%%%%%%%%%%%%%%%%%%%%%%
\subsection{Translation of Geometry between \geant and \mitsuba}
\label{sec:geo_translation}
It is necessary to transform the geometry of a RICH detector from  
\geant GDML format into a format suitable for rendering in \mitsuba. First, the geometry is visualised within \freecad~\cite{freecad}. The geometry was simplified by removing the detector casing and related components including windows, the remaining mirrors and detector plane were grouped into OBJ files, and incorporated into an XML format file which is readable by \mitsuba. Each stage of the process introduces its own set of considerations, and the techniques and strategies adopted to address these are discussed in the following. 

\freecad is an open-source 3D computer-aided design (CAD) platform and is frequently used for creating, manipulating, and analysing 3D models. However, \freecad does not naively support the GDML format so the open-source \freecad plugin \texttt{CAD\_GDML}\cite{cad_gdml} was used. This plugin serves as an effective intermediary, facilitating the interaction between the GDML input file and \freecad's visualisation interface. 

\begin{figure}
    \centering
    \includegraphics[width=0.3\textwidth]{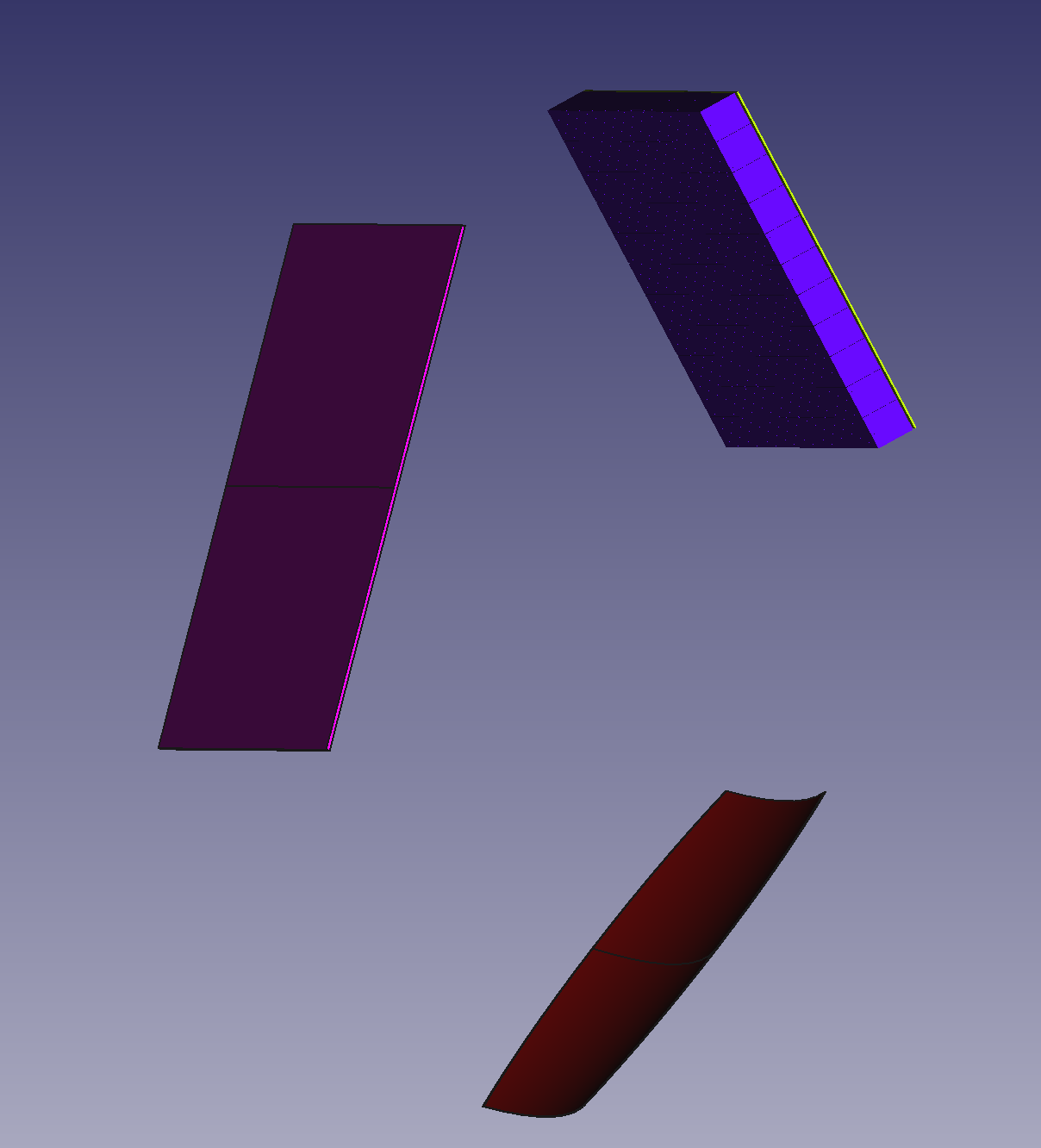}
    \caption{Simplified RICH geometry visualised in \freecad. Detector components are visible, including a spherical mirror positioned at the bottom right, a flat mirror on the left, and a detector at the top right corner in the scene.}
    \label{fig:-FreeCAD-model}
\end{figure}

With this a simplified model of the RICH geometry was created as shown in Figure~\ref{fig:-FreeCAD-model}.  

The models corresponding to the aforementioned components are exported as a pair of distinct OBJ files: the first incorporates both mirrors as a unified entity, while the second represents the detector. This strategy is adopted for two primary reasons: initially, treating the mirrors as a unified entity negates the requirement for subsequent angular modifications; furthermore, the mirrors are fabricated from identical material, facilitating the usage of a shared Bidirectional Scattering Distribution Function (BSDF). The BSDF, one of the \mitsuba plugins, characterises the surface materials of the objects and simulates the interaction of light with different surfaces. Conversely, the detector elements, composed of a different material, are best used when separated from the mirrors.

For the transformation of objects from \freecad to OBJ files, an embedded mesh generation method \texttt{Mefisto} is used to give a high resolution model, without compromising on approximations. The operation of this method is governed by a singular parameter, the maximum edge length. This parameter limits the edge lengths of individual triangles or polygons generated during the meshing phase. In the present instance, a maximum edge length of 5 mm is selected. 

Upon integrating the OBJ files into the XML file for rendering in \mitsuba, it is important that the coordinate systems, interrelated positions and scale of the models remain comparable to those in the GDML file. Consequently, additional model transformations within the XML file are not necessary.

We note that other options for simplification of the geometry are possible---for instance exporting only a single sub-detector to a \geant parallel world, followed by exporting of this world to an individual GDML file adequately removes many steps. Finally, the use of other tools, such as \texttt{pyg4ometry}~\cite{WALKER2022108228} allow the conversion to OBJ formats in a single script. Implementation of these options is beyond the scope of this paper.

%%%%%%%%%%%%%%%%%%%%%%%%%%%%%%%%%%%%%%%%%%%%%%%%%%%
\subsection{Photon Emitter}
\label{sec:custom_emitter}
In \mitsuba, the \texttt{emitter} plugins are used to initiate light rays. The spot light emitter (\texttt{spot}) is the closest available emitter that represents the behaviour of a photon. This plugin produces a conical light with linear falloff, governed by the \texttt{cutoff\_angle} parameter that restricts the light emission within a specified angular range. Ideally, for photon propagation, it should be possible to represent a single photon by sufficiently minimising the \texttt{cutoff\_angle} of a solitary (\texttt{spot}) emitter.

However, as the spot light emitter emits a cone shape light, regardless of the value of the parameter {\texttt{cutoff\_angle}}, the size of the spot increases and its brightness decreases as the light path length increases, insufficient for the representation of single photons. Furthermore, as each spot emitter can only represent a single photon, many are needed to represent the emission of Cherenkov radiation over a charged particle's trajectory within the radiator. The \drjit compiler creates a separate C++ object for each emitter. As such large simulations containing millions of photons can easily surpass a GPU's global memory capacity, this approach can result in slow compilation speeds and rendering failures. This representation of photons and the significant memory usage associated with it meant using a single (\texttt{spot}) emitter to represent one photon is not a viable option. 

To rectify these issues the \texttt{photon\_emitter} has been developed as a custom emitter by modifying the (\texttt{spot}) emitter. This new emitter initiates multiple light rays by taking a vector input of initial positions and momenta read from a binary format file or {\sc Numpy} array. Furthermore we set the \texttt{local\_dir} parameter in the \texttt{sample\_ray} function to a constant thus fixing the emitter's orientation. The \texttt{falloff\_curve} parameter in the \texttt{photon\_emitter} is set to unity to ensure constant intensity. Given that only one light ray is required per emission, the concept of the \texttt{cutoff\_angle} parameter becomes irrelevant and has been removed. As the \texttt{photon\_emitter} can initiate every photon in a simulation with a single instantiation, the memory footprint of kernel generated by \drjit is reduced substantially, from gigabytes to kilobytes. Moreover, for large simulations containing millions of photons, the compilation time was reduced from hours to seconds. 

%%%%%%%%%%%%%%%%%%%%%%%%%%%%%%%%%%%%%%%%%%%%%%%%%%%
%%%%%%%%%%%%%%%%%%%%%%%%%%%%%%%%%%%%%%%%%%%%%%%%%%%
\subsection{Modelling of Quantum Efficiency}
\label{sec:quantum_eff}

\begin{figure}[!htp]
\centering
    \includegraphics[width=1\columnwidth]{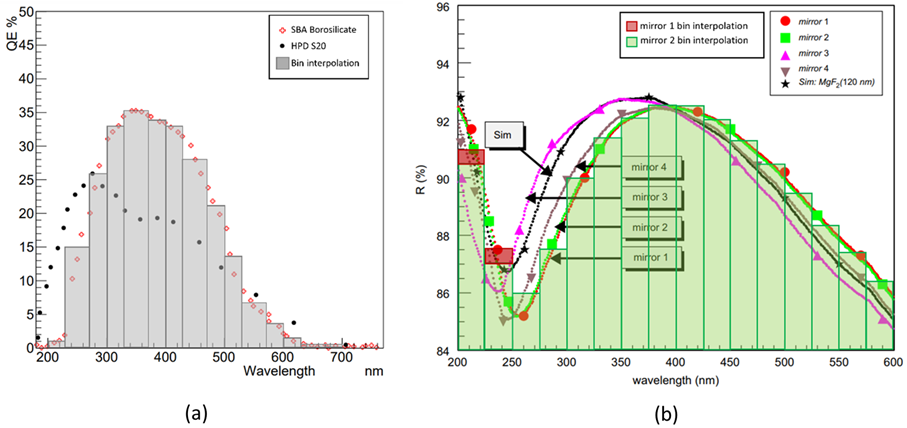}
    \caption{\small Efficiencies of the detector and mirrors against the wavelength of incident light, from Ref~\cite{LHCB_TDR_014}. (a) The quantum efficiency of Hybrid Photon Detectors (in black), of MaPMT sensors (in red) and the interpolation created for the simulation in \mitsuba. (b) The reflectivity of the four mirrors segments composing the spherical mirrors against wavelength. The red and green bins show the interpolation of mirrors $1$ \& $2$ implemented in the simulation. }
    \label{fig:efficiencies}
\end{figure}

The efficiency of detection of Cherenkov photons relies on the intrinsic efficiency of the photon detectors in question and must be accounted for. In the case of the simplified RICH detector, one must consider the efficiency of the Multi-Anode Photo Multiplier Tube (MaPMT) detectors \cite{Hamamatsu} and the reflectivity of the mirrors, which are wavelength dependent. Furthermore, as computational expense is dependent on the number of photons, it is advantageous to apply any quantum efficiency and reflectivity effects prior to propagation and discard the associated photons. This is justified, as any discarded photons would not be detected. Figure \ref{fig:efficiencies}(a) describes the quantum efficiency of the detector as enumerated in Ref.~\cite{LHCB_TDR_014}, which peaks at $35\%$ around $450~\si{nm}$, due to the Borosilicate window's refractive index variations in the $200$ to $1000~\si{nm}$ region. Implementation of the efficiency was performed by transforming the measured efficiencies into histograms of $33~\si{nm}$ bin width. The reflectivity of the mirrors, given in Figure \ref{fig:efficiencies}(b) vary between $94\%$ and  $89\%$ between $200$ and $600~\si{nm}$. Subsequent estimation of the reflectivity was performed in a similar fashion to the efficiency, utilising bins of $25~\si{nm}$. For each photon, with its specific wavelength, an accept-reject algorithm was used to model the three efficiencies before propagation. As described in section \ref{sec:mitsuba_summ}, this approach is only valid when using the \texttt{mono} variants, which are inherently faster than the variants that render multiple wavelengths. 

%%%%%%%%%%%%%%%%%%%%%%%%%%%%%%%%%%%%%%%%%%%%%%%%%%%
\subsection{Translation to global coordinate systems}
\label{sec:adjustments}

In \mitsuba, detection is achieved using the \texttt{camera} and \texttt{film} plugins to sample the propagated rays. The detector plane in the simplified RICH geometry measured $620 \times 1320 ~\si{mm}$, which defined the size of the film. The \texttt{camera} was placed at a distance of $478.69 ~\si{mm}$, facing the centre of the detector. The field of view was empirically set to $116.38^{\circ}$ to match the size of the detector. These values where determined empirically by matching the radii of the Cherenkov rings. The bitmap output was translated into a {\sc Numpy} detection matrix. The \texttt{mono} variants created a $620 \times 1320$ two-dimensional array of intensities at each pixel ranging between $0$ and $255$. A temporary method was used to set an empirical threshold of intensity of $45$ in local units to keep the same number of hits as rays propagated, only accepting values above this threshold. This solution is not viable but enabled checking the implementation of the geometry. The position of each cell passing this threshold was recorded as a photons hit; these were 2D coordinates relative to the bottom left corner of the detector plane.

\begin{figure}[!htp]
\centering
    \includegraphics[width=1\columnwidth]{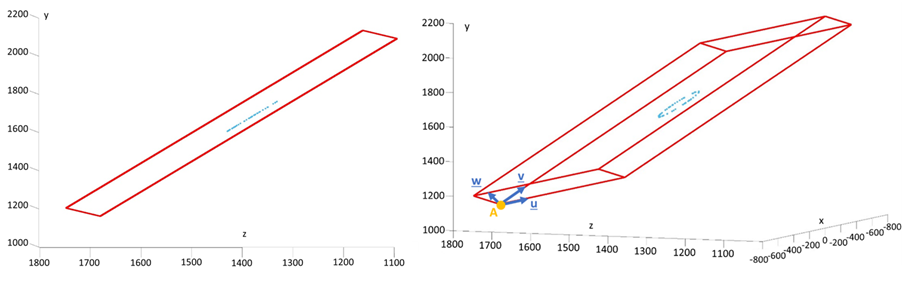}
    \caption{\small Plots of the output from \geant simulation of the simplified RICH geometry. In red is the detector volume. On the left, a slice in the y-z plane shows the hits inside of the volume. On the right, a 3D view showing the unit vectors defined on the surface of the detector and point A, the bottom left corner.}
    \label{fig:3D_G4}
\end{figure}

Detected photons in \geant are referred to as \texttt{hits} and are represented as $x,y,z$ coordinates relative to an origin defined in the geometry in this case, the centre of the RICH detector. In Figure~\ref{fig:3D_G4} we see the hits as recorded in \geant relative to the detector volume. There is an offset due to the simplification of geometry, in \geant detection occurs at the back side of the MaPMTs whereas for the simplified geometry used in \mitsuba detection occurs at the front side of the detector volume shown in red in Figure~\ref{fig:3D_G4}. The hits in \geant were projected onto the surface of the detector volume creating two-dimensional coordinates allowing for direct comparison to \mitsuba, as illustrated in Figure \ref{fig:3D_G4}. This issue is easily remedied by taking the final position of the rays as global coordinates, but requires a change to the underlying \mitsuba output.

%%%%%%%%%%%%%%%%%%%%%%%%%%%%%%%%%%%%%%%%%%%%%%%%%%%
%%%%%%%%%%%%%%%%%%%%%%%%%%%%%%%%%%%%%%%%%%%%%%%%%%%
\section{Results}
\label{sec:results}
%%%%%%%%%%%%%%%%%%%%%%%%%%%%%%%%%%%%%%%%%%%%%%%%%%%
\subsection{Comparison of Cherenkov Rings}
\label{sec:result_rings}

Simulations of Cherenkov photons using the simplified RICH geometry were first performed in \geant using a $100 ~\si{GeV}$ $\mu^{+}$ particle. The origin and momenta of the emitted Cherenkov photons were output in binary format and used as input to \mitsuba, following the described pipeline.

\begin{figure}[!htp]
\centering
    \includegraphics[width=1\columnwidth]{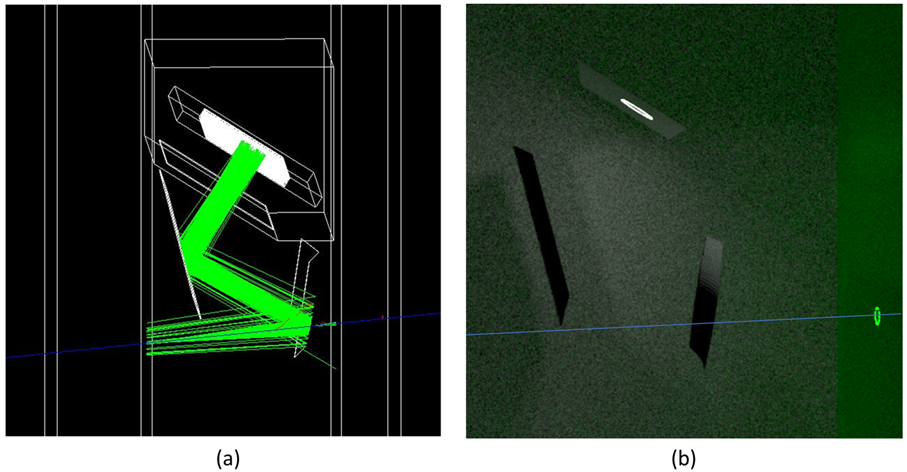}
    \caption{\small (a) The scene in \geant showing the path of the $\mu^{+}$ (in blue) and photons (in green), and their interactions with the subdetector elements. Photons can be spotted behind the spherical mirror due to emission after passing the mirror. (b) The corresponding scene in \mitsuba shows the trajectory of the charged particle (in blue). A green wall was added on the left to let photons emitted after the spherical mirror diffuse. Rings of light can be observed on the detector and on the wall.}
    \label{fig:rays_g4_m3}
\end{figure}

Figure \ref{fig:rays_g4_m3}(a) shows the path of the $\mu^{+}$ and of the emitted Cherenkov photons in \geant. At the bottom are the origins of the photons. At the bottom on the right, a smaller green cone can be seen emerging from the particle's path behind the spherical mirror due to emission past the first mirror. 
Figure \ref{fig:rays_g4_m3}(b) shows the corresponding scene in \mitsuba. For the sake of visualisation, spot emitters were used to show the end of the trajectory of the hits. At the top, a ring of light can be seen. A green wall was added on the right of the scene to show a second ring of light in the bottom right corner, behind the spherical mirror due to photons emitted by the $\mu^{+}$ after passing through it.

Both scenes demonstrate identical light behaviours. The ring of light forming behind the spherical mirror supported the good calibration of the objects in \mitsuba. 

\begin{figure}[!htp]
\centering
    \includegraphics[width=1\columnwidth]{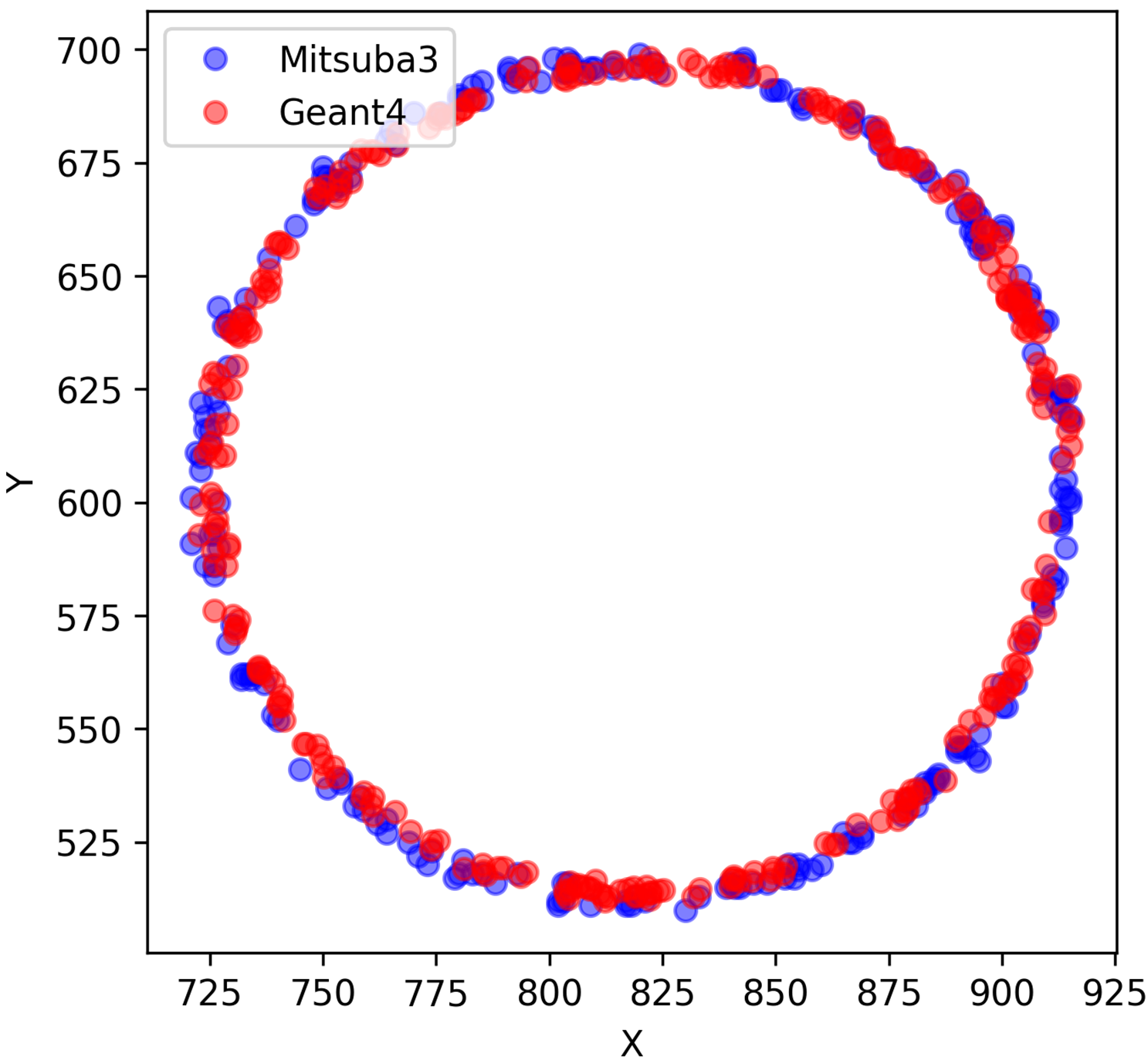}
    \caption{\small Position in pixels of detector hits in \geant (red) and \mitsuba (blue).}
    \label{fig:output_g_m}
\end{figure}

Hits were simulated in both software including the quantum efficiency and reflectivity and are shown in Figure \ref{fig:output_g_m}. The Cherenkov rings returned by \geant and \mitsuba are comparable, exhibiting similar radii, however, the absolute positions differ. Results are discussed later in section \ref{sec:discussion}.

%%%%%%%%%%%%%%%%%%%%%%%%%%%%%%%%%%%%%%%%%%%%%%%%%%%
\subsection{Timing}
\label{sec:timing}
%%%%%%%%%%%%%%%%%%%%%%%%%%%%%%%%%%%%%%%%%%%%%%%%%%%
Both the wavelength treatment and parallelisation implementation are selected at run-time using variants. This study tested and compared four variants, \texttt{llvm\_mono},\\ \texttt{llvm\_rgb}, \texttt{cuda\_mono}, and \texttt{cuda\_rgb}, which are described in Section~\ref{sec:variants}. These variants were chosen to test both relative speed-up against \geant for CPU and GPU propagation; as well as for wavelength treatment as these will become important in the future. 

The CPU used in testing was an Intel(R) Xeon(R) Silver 4210R 2.40GHz with 20 cores, each timing simulation was executed with 1,2,4,8,16 and 20 threads and repeated three times to eliminate outliers. Testing on GPU was performed on a NVIDIA Tesla T4 with 2560 threads, a base clock of 585 MHz (1590 MHz boost clock) and 40 Ray Tracing (RT) cores.

\mitsuba has in built timing reports which are generated at run-time for a variety of rendering stages. Timings were confirmed using the NVIDIA {\sc NSight} tools \cite{nvidia_nsight}. Through investigation of the timing reports, it was determined that there were three areas of computation expense, parsing of the XML file, initialisation of the kernel which includes the \drjit compilation and rendering of the scene.

%%%%%%%%%%%%%%%%%%%%%%%%%%%%%%%%%%%%%%%%%%%%%%%%%%%
\section{Discussion}
\label{sec:discussion}

In Figure~\ref{fig:output_g_m}, the overlap of the rings showed that the centre of both fell in the same position. This confirms the comparable behavior of both \geant and \mitsuba and the accurate translation of the geometry. The initial coordinates of the photons were directly exported from \geant to \mitsuba; therefore, the similarity of the ring sizes verifies that the coordinate system in \mitsuba corresponds to that of \geant. Finally, corroborating again the correct geometry, the comparable radii are the key result of the simulation. The Cherenkov angle is calculated from the radius of the ring and thus the speed of the charged particle can be identified. Both rings will give the same charged particle's speed and demonstrate \mitsuba's potential for future implementation.

Two issues arose from Figure \ref{fig:output_g_m}. The exact position of the hits did not correspond, and the radius of the ring in \mitsuba seemed slightly larger than that of \geant. The clash in the position of the hits came from the random filtering of the photons described in section \ref{sec:quantum_eff} and the method used to identify hits on the detector described in section \ref{sec:geo_translation}. The threshold was set to $45$ to identify as many rays as emitted after filtering as possible. Some pixels in the detector were brighter than the implemented threshold due to signal leakage to neighboring pixels. Some other photons fell between pixels and gave a maximum pixel intensity value of less the threshold. The radii mismatch is due to the discussed scale mismatch. Photons in \geant and \mitsuba did not effectively hit the same surface (see Figure \ref{fig:3D_G4}) and the projection of the hits onto the surface introduced errors. The recording of final positions in \mitsuba was made through a camera with a field of view determined with approximations. A variation of a tenth of degree in the field of view could offset the position of the hits. Finally, the pixel hits were detected, and not their exact positions, providing an accuracy of $1 ~\si{mm}$ at best - the size of a pixel on the film.  These issues can be mitigated by the output of individual global hit positions as opposed to direct detector response.

\begin{figure}
    \centering
    \includegraphics[width=1\columnwidth]{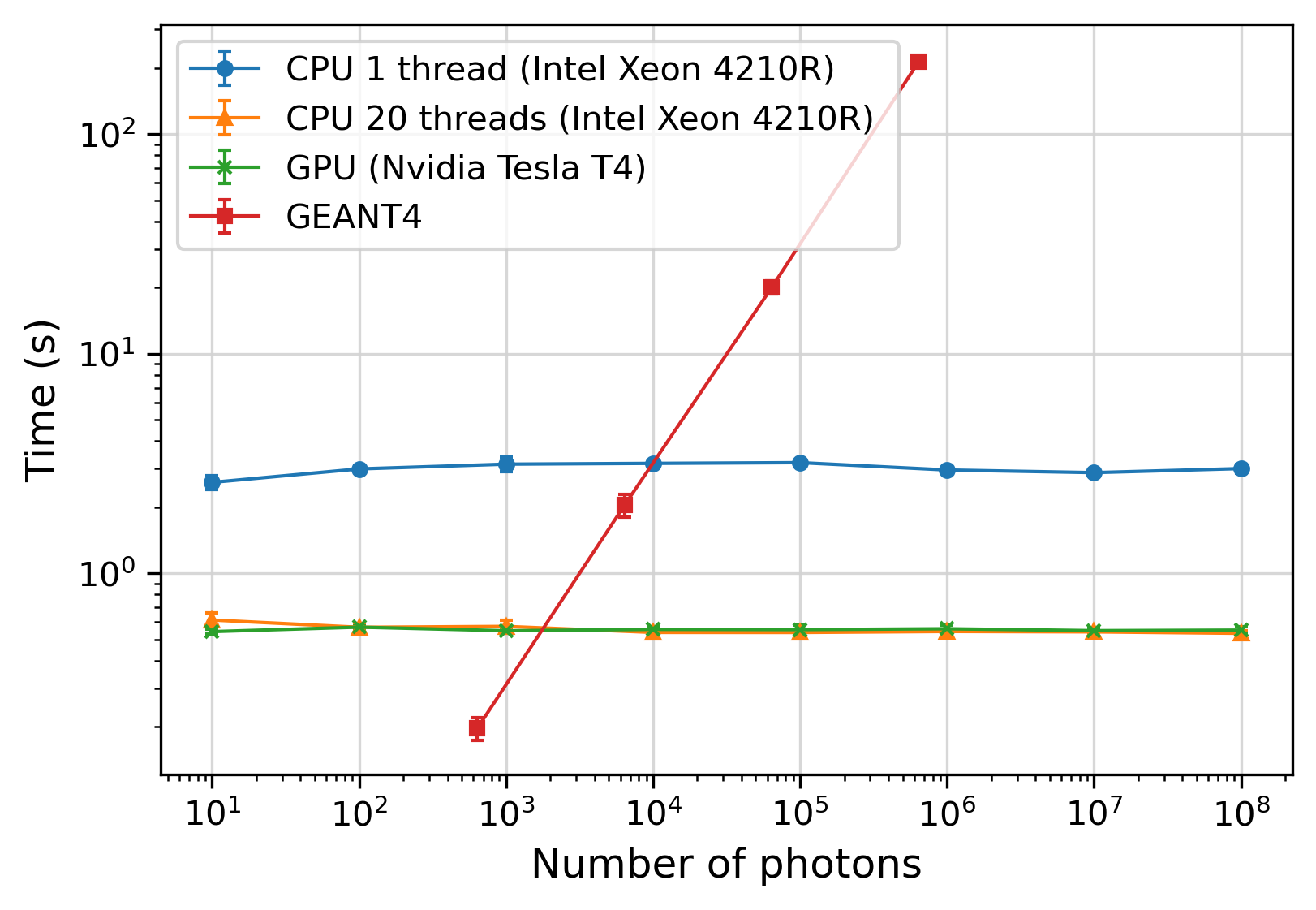}
    \caption{\small Scaling profile of the \geant simulation compared to that of \mitsuba, which was executed on three different computational resources: an Intel Xeon 4210R CPU with 1 and 20 threads, and an NVIDIA Tesla T4 GPU. In contrast, \geant was run on a single thread on the Intel Xeon 4210R CPU.}
    \label{fig:scaling_profile}
\end{figure}

The \geant simulation used for comparison with \\\mitsuba was run on a single thread on the CPU, and hence, was not parallelised. The time taken to propagate photons was measured for simulations that produced $600$ to $6\times10^5$ photons, in steps of factors of 10. The scaling profile of \mitsuba on different architectures and that of \geant are presented in Figure~\ref{fig:scaling_profile}. The first observation is that, contrary to the \geant simulation, the \mitsuba implementation does not scale with the number of photons. This indicates that there is a step in the rendering process that is the same over all of the different variations and which dominates the total time. This is known to be the production of the images but can be improved by writing the global coordinates of the photons, directly bypassing this step. Furthermore, this is independent of the architecture since both the CPU and the GPU exhibit this behaviour.

The second observation is that \mitsuba outperforms \geant when simulating $\geq~2000$ photons on the GPU or on the CPU at 20 cores, and when propagating $\geq~10^4$ photons on the CPU with a single thread. Beyond these thresholds, \mitsuba then becomes increasingly faster than \geant. For example, when considering the maximum number of photons simulated by \geant, i.e. $6\times10^5$, \mitsuba is approximately 70 times faster on the CPU at a single thread and 400 times faster on both the GPU and CPU at 20 threads. Another way of quantifying the performance of the two pipelines is by calculating the photo propagation rate i.e. the number of photons propagated per second. The \mitsuba framework is capable of rendering about $1.85\times10^8$ photons per second on both the GPU and on the CPU at 20 threads, and $3.3\times10^7$ photons per second on the single-threaded CPU. By contrast, the \geant simulation is capable of propagating up to $3200$ photons per second. Therefore, for a workload greater than the threshold values, offloading the propagation of Cherenkov photons to the \mitsuba pipeline will be significantly more efficient than simulating them using \geant.

\begin{figure}
    \centering
    \includegraphics[width=\columnwidth]{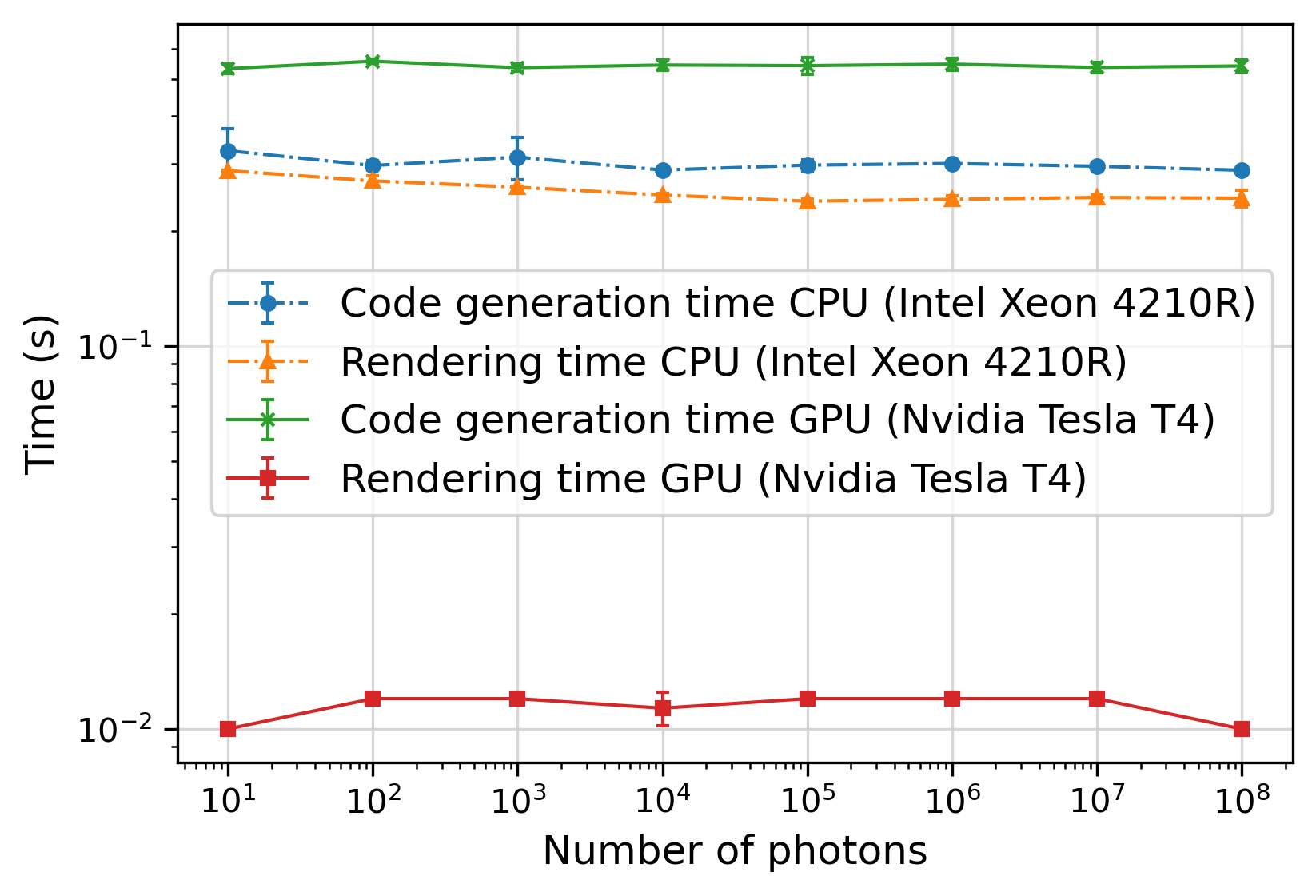}
    \caption{Breakdown of the photon propagation time into the code generation and rendering time as a function of the number of photons for both the Intel Xeon 4210R CPU at 20 threads and the NVIDIA Tesla T4.}
    \label{fig:codegen_rendering}
\end{figure}

Finally, the third notable feature of this plot is the similar performance between the GPU and the CPU at 20 threads. This is odd behavior at first sight due to the fact that the 
NVIDIA GPU has a greater number of cores compared to the Intel CPU and that, in general, GPUs are better suited for ray tracing. Breaking down the simulation time into the code generation time and rendering time separately, as shown in Figure~\ref{fig:codegen_rendering}, reveals that when running \mitsuba on the GPU the code generation time significantly outweighs the rendering time. On the other hand, when running \mitsuba on the CPU at 20 cores both times are comparable. Isolating the rendering time, the GPU is $22$ times faster than the CPU. Therefore, optimising the code generation when using the GPU can significantly improve \mitsuba's performance, thereby further reducing the threshold number of photons required to outperform \geant on that architecture.

\section{Conclusion and future work}\label{sec13}
\label{sec:conclusion}

 A successful prototype workflow offloading Cherenkov photon propagation from \geant to \mitsuba has been implemented. A light source analogue to a photon consisting of a single ray with infinitesimal width, in a single direction, and without decaying intensity was successfully created in \mitsuba. The intrinsic efficiencies of each object contained in the simplified LHCb RICH detector were reproduced and the detection of photons propagated with \mitsuba was successful. The comparison of the outputs confirmed the correct implementation of a simplified RICH geometry and the accurate propagation of the photons within that geometry. The timing study showed that, up to $10^8$ photons, the simulation time in \mitsuba does not scale with the number of photons. Above $10^4$ photons, \mitsuba is significantly faster than \geant. The simulation in \mitsuba is not yet ready to be integrated into \geant but, with further development, this is expected to be feasible and to yield significant performance gains.

\mitsuba proved its potential for physics experiments involving light propagation. However, the graphical results are not yet accurate enough for the substitution of \geant.
 The transition to memory-resident objects is essential. Such a change will allow the output method to return the global coordinates of the hits. In a similar fashion to what was done for the \texttt{photon\_emitter} in the source code, a new BSDF class should be constructed for the detector. BSDFs are called every time a ray intersects with a new interface. The BSDF should include a function that writes the position of the ray in a list when it interacts with the detector. This would avoid deformation of the output either due to the leakage issue of hits being created in multiple pixels or due to inaccuracies in the camera's field of view. The scaling mismatch described in Section \ref{sec:result_rings} would be solved, hits from \geant and \mitsuba could be directly compared, and statistical analysis could be conducted to better judge the potential of \mitsuba for the simplified RICH detector. Later, hits from \mitsuba could be directly fed back into the global \geant simulation, allowing further processing.

The photon filtering for quantum efficiency and reflectivity should be improved in one of two ways. The first method is to implement the filtering within the overarching \texttt{C++} simulation to automate the application to each new set of photons produced by \geant. Another way is to design a wavelength-dependent BSDF with a function to kill or propagate rays as they are traced and interact with the different objects in \mitsuba. This could involve using colour-dependent variants or attaching a value for the wavelength to each ray traced---possibly in place of the intensity. Both methods could be studied by isolating the effect of each filtering in \geant and \mitsuba and comparing the fraction of photons detected.

Finally, the removal of all intermediate steps represents a complete \texttt{C++} pipeline of passing of photons from \geant to \mitsuba, their filtering, their loading into the \texttt{photon\_emitter}, their propagation, their detection and the loading of the hits back to \geant. 

In addition to the simulation of Cherenkov photons in gaseous detectors, which was the focus of the work presented here, the use of \mitsuba is also expected to yield similar improvements in other simulations involving optical photons. This includes the simulation of Cherenkov photons in liquids, such as water Cherenkov detectors, and other light sources in liquids including scintillation light in liquid noble gas detectors.

\section*{Acknowledgments}

The authors would like to thank S. Easo and Y. Li for discussions related to the simplified RICH Geometry. We thank Nicolas Roussel from \mitsuba for helpful discussions.
ACSD and MG acknowledge funding from UKRI-STFC under grant reference ST/W000601/1. % CG
KE acknowledges funding from UKRI-STFC under grant reference ST/V002546/1 %SWIFT-HEP
and from the Royal Society (United Kingdom) under grant agreements DH160214 and RGF/EA/201014. 

\bibliographystyle{LHCb}
\bibliography{sn-bibliography}% common bib file

\ifx\mcitethebibliography\mciteundefinedmacro
\PackageError{LHCb.bst}{mciteplus.sty has not been loaded}
{This bibstyle requires the use of the mciteplus package.}\fi
\providecommand{\href}[2]{#2}
\begin{mcitethebibliography}{10}
\mciteSetBstSublistMode{n}
\mciteSetBstMaxWidthForm{subitem}{\alph{mcitesubitemcount})}
\mciteSetBstSublistLabelBeginEnd{\mcitemaxwidthsubitemform\space}
{\relax}{\relax}

\bibitem{Mitsuba3}
W.~Jakob {\em et~al.}, \ifthenelse{\boolean{articletitles}}{\emph{Mitsuba 3
  renderer}, }{} 2022.
\newblock https://mitsuba-renderer.org\relax
\mciteBstWouldAddEndPuncttrue
\mciteSetBstMidEndSepPunct{\mcitedefaultmidpunct}
{\mcitedefaultendpunct}{\mcitedefaultseppunct}\relax
\EndOfBibitem
\bibitem{Agostinelli:2002hh}
Geant4 collaboration, S.~Agostinelli {\em et~al.},
  \ifthenelse{\boolean{articletitles}}{\emph{{Geant4: A simulation toolkit}},
  }{}\href{https://doi.org/10.1016/S0168-9002(03)01368-8}{Nucl.\ Instrum.\
  Meth.\  \textbf{A506} (2003) 250}\relax
\mciteBstWouldAddEndPuncttrue
\mciteSetBstMidEndSepPunct{\mcitedefaultmidpunct}
{\mcitedefaultendpunct}{\mcitedefaultseppunct}\relax
\EndOfBibitem
\bibitem{Allison:2006ve}
Geant4 collaboration, J.~Allison {\em et~al.},
  \ifthenelse{\boolean{articletitles}}{\emph{{Geant4 developments and
  applications}}, }{}\href{https://doi.org/10.1109/TNS.2006.869826}{IEEE
  Trans.\ Nucl.\ Sci.\  \textbf{53} (2006) 270}\relax
\mciteBstWouldAddEndPuncttrue
\mciteSetBstMidEndSepPunct{\mcitedefaultmidpunct}
{\mcitedefaultendpunct}{\mcitedefaultseppunct}\relax
\EndOfBibitem
\bibitem{Li:2023ocy}
Y.~Li {\em et~al.}, \ifthenelse{\boolean{articletitles}}{\emph{{GPU-based
  optical photon simulation for the LHCb RICH 1 Detector}},
  }{}\href{http://arxiv.org/abs/2307.10823}{{\normalfont\ttfamily
  arXiv:2307.10823}}\relax
\mciteBstWouldAddEndPuncttrue
\mciteSetBstMidEndSepPunct{\mcitedefaultmidpunct}
{\mcitedefaultendpunct}{\mcitedefaultseppunct}\relax
\EndOfBibitem
\bibitem{Jakob2020DrJit}
W.~Jakob, S.~Speierer, N.~Roussel, and D.~Vicini,
  \ifthenelse{\boolean{articletitles}}{\emph{Dr.jit: A just-in-time compiler
  for differentiable rendering},
  }{}\href{https://doi.org/10.1145/3528223.3530099}{Transactions on Graphics
  (Proceedings of SIGGRAPH) \textbf{41} (2022) }\relax
\mciteBstWouldAddEndPuncttrue
\mciteSetBstMidEndSepPunct{\mcitedefaultmidpunct}
{\mcitedefaultendpunct}{\mcitedefaultseppunct}\relax
\EndOfBibitem
\bibitem{cuda}
NVIDIA, P.~Vingelmann, and F.~H.~P. Fitzek,
  \ifthenelse{\boolean{articletitles}}{\emph{Cuda, release: 10.2.89}, }{}
  2020\relax
\mciteBstWouldAddEndPuncttrue
\mciteSetBstMidEndSepPunct{\mcitedefaultmidpunct}
{\mcitedefaultendpunct}{\mcitedefaultseppunct}\relax
\EndOfBibitem
\bibitem{optix}
S.~G. Parker {\em et~al.}, \ifthenelse{\boolean{articletitles}}{\emph{{GPU Ray
  Tracing}}, }{}\href{https://doi.org/10.1145/2447976.2447997}{Communications
  of the ACM \textbf{56} (2013) 93}\relax
\mciteBstWouldAddEndPuncttrue
\mciteSetBstMidEndSepPunct{\mcitedefaultmidpunct}
{\mcitedefaultendpunct}{\mcitedefaultseppunct}\relax
\EndOfBibitem
\bibitem{LLVM:CGO04}
C.~Lattner and V.~Adve, \ifthenelse{\boolean{articletitles}}{\emph{{LLVM}: A
  compilation framework for lifelong program analysis and transformation}, }{}
  (San Jose, CA, USA), 75--88, 2004\relax
\mciteBstWouldAddEndPuncttrue
\mciteSetBstMidEndSepPunct{\mcitedefaultmidpunct}
{\mcitedefaultendpunct}{\mcitedefaultseppunct}\relax
\EndOfBibitem
\bibitem{Wald:2014:EKF:2601097.2601199}
I.~Wald {\em et~al.}, \ifthenelse{\boolean{articletitles}}{\emph{Embree: A
  kernel framework for efficient cpu ray tracing},
  }{}\href{https://doi.org/10.1145/2601097.2601199}{ACM Trans.\ Graph.\
  \textbf{33} (2014) 143:1}\relax
\mciteBstWouldAddEndPuncttrue
\mciteSetBstMidEndSepPunct{\mcitedefaultmidpunct}
{\mcitedefaultendpunct}{\mcitedefaultseppunct}\relax
\EndOfBibitem
\bibitem{freecad}
J.~Riegel, W.~Mayer, and Y.~van Havre,
  \ifthenelse{\boolean{articletitles}}{\emph{Freecad: Your own 3d parametric
  modeler}, }{} \url{https://www.freecad.org/}, 2021\relax
\mciteBstWouldAddEndPuncttrue
\mciteSetBstMidEndSepPunct{\mcitedefaultmidpunct}
{\mcitedefaultendpunct}{\mcitedefaultseppunct}\relax
\EndOfBibitem
\bibitem{cad_gdml}
K.~Sloan, \ifthenelse{\boolean{articletitles}}{\emph{Cad-gdml project}, }{}
  \url{http://cad-gdml.in2p3.fr/}, 2020\relax
\mciteBstWouldAddEndPuncttrue
\mciteSetBstMidEndSepPunct{\mcitedefaultmidpunct}
{\mcitedefaultendpunct}{\mcitedefaultseppunct}\relax
\EndOfBibitem
\bibitem{WALKER2022108228}
S.~D. Walker {\em et~al.},
  \ifthenelse{\boolean{articletitles}}{\emph{Pyg4ometry: A python library for
  the creation of monte carlo radiation transport physical geometries},
  }{}\href{https://doi.org/https://doi.org/10.1016/j.cpc.2021.108228}{Computer
  Physics Communications \textbf{272} (2022) 108228}\relax
\mciteBstWouldAddEndPuncttrue
\mciteSetBstMidEndSepPunct{\mcitedefaultmidpunct}
{\mcitedefaultendpunct}{\mcitedefaultseppunct}\relax
\EndOfBibitem
\bibitem{Hamamatsu}
Hamamatsu, \ifthenelse{\boolean{articletitles}}{\emph{{Multi Anode
  photomultiplier tube R11265U-100-M4}}, }{}
  \url{https://www.hamamatsu.com/us/en/product/optical-sensors/pmt/pmt_tube-alone/metal-package-type/R11265U-100-M4.html},
  2022\relax
\mciteBstWouldAddEndPuncttrue
\mciteSetBstMidEndSepPunct{\mcitedefaultmidpunct}
{\mcitedefaultendpunct}{\mcitedefaultseppunct}\relax
\EndOfBibitem
\bibitem{LHCB_TDR_014}
{LHCb Collaboration}, \ifthenelse{\boolean{articletitles}}{\emph{{LHCb PID
  Upgrade Technical Design Report}}}{} , 2013\relax
\mciteBstWouldAddEndPuncttrue
\mciteSetBstMidEndSepPunct{\mcitedefaultmidpunct}
{\mcitedefaultendpunct}{\mcitedefaultseppunct}\relax
\EndOfBibitem
\bibitem{nvidia_nsight}
{NVIDIA Corporation}, \ifthenelse{\boolean{articletitles}}{\emph{Nvidia nsight
  systems}, }{} \url{https://developer.nvidia.com/nsight-systems}, 2023\relax
\mciteBstWouldAddEndPuncttrue
\mciteSetBstMidEndSepPunct{\mcitedefaultmidpunct}
{\mcitedefaultendpunct}{\mcitedefaultseppunct}\relax
\EndOfBibitem
\end{mcitethebibliography}

\end{document}